\documentclass{emulateapj}

\usepackage{graphics}
\usepackage{epsfig}
\usepackage{natbib}
\shorttitle{Auroral/Dayglow Observations of HD209458b}
\shortauthors{France et al.}

\begin{document}


\title{Searching for Far-Ultraviolet Auroral/Dayglow Emission from HD209458b.\altaffilmark{*}}


\author{Kevin France\altaffilmark{1}, John T. Stocke} 
\affil{Center for Astrophysics and Space Astronomy, 389 UCB, University of Colorado, 
Boulder, CO 80309}

\author{Hao Yang, Jeffrey L. Linsky} 

\affil{JILA, University of Colorado, Boulder, CO 80309}

\author{Brian C. Wolven }
\affil{Applied Physics Laboratory, Johns Hopkins University, 11100 Johns Hopkins Rd, Laurel, MD 20723}

\author{Cynthia S. Froning, James C. Green, and Steven N. Osterman} 

\affil{Center for Astrophysics and Space Astronomy, 389 UCB, University of Colorado, 
Boulder, CO 80309}



\altaffiltext{*}{Based on observations made with the NASA/ESA $Hubble$~S$pace$~$Telescope$, obtained from the data archive at the Space Telescope Science Institute. STScI is operated by the Association of Universities for Research in Astronomy, Inc. under NASA contract NAS 5-26555.}

\altaffiltext{1}{kevin.france@colorado.edu}




\begin{abstract}
We present recent observations from the $HST$-Cosmic Origins Spectrograph aimed at characterizing the auroral emission from the extrasolar planet HD209458b.  We obtained medium-resolution ($R$~$\sim$~20,000)
far-ultraviolet (1150~--~1700 \AA) spectra at both the Phase 0.25 and Phase 0.75 quadrature positions as well as a stellar baseline measurement at secondary eclipse.  This analysis includes a catalog of stellar emission lines and a star-subtracted spectrum of the planet.  We present an emission model for planetary H$_{2}$ emission, and compare this model to the planetary spectrum.  No unambiguously identifiable atomic or molecular features are detected, and upper limits are presented for auroral/dayglow line strengths.  An orbital velocity cross-correlation analysis finds a statistically significant (3.8~$\sigma$) 
feature at +15	($\pm$ 20) km s$^{-1}$ in the rest frame of the planet, at $\lambda$1582~\AA.  This feature is consistent with emission from H$_{2}$ $B$~--~$X$ (2~--~9) P(4) ($\lambda_{rest}$~=~1581.11~\AA), however the physical mechanism required to excite this transition is unclear.  We compare limits on relative line strengths seen in the exoplanet spectrum with models of ultraviolet fluorescence to constrain the atmospheric column density of neutral hydrogen between the star and the planetary surface.
These results support models of short period extrasolar giant planets with weak magnetic fields and extended atomic atmospheres.
\end{abstract}



\clearpage


\section{Introduction}

The preceding two decades have witnessed an explosion in both the number of known extrasolar
planets and the number of astronomers studying them.  There has been over an order of magnitude increase in the number of detected exoplanets in the last ten years alone, now totaling over 400.  Among these, transiting planets have proven to be the most 
useful for constraining the physical and chemical conditions of the planets and their atmospheres.  
HD209458b is a short-period ($a$~=~0.046 AU; Henry et al. 2000) transiting gas giant, with a mass of $\sim$~0.7~$M_{J}$~\citep{laughlin05} and a density ~$\sim$~25\% that of Jupiter.  HD209458b orbits a solar-type (G0V) star, which is located at a distance of 47 pc.  
It was the first extrasolar planet observed in transit~\citep{henry00,charbonneau00}, has a well determined orbit from radial velocity studies~\citep{mazeh00}, and was the first object whose atmosphere was probed by absorption line spectroscopy (Charbonneau et al. 2002; and see Sing et al. 2008a,b for additional analysis).~\nocite{charbonneau02,henry00,sing08a,sing08b}  

HD209458b was also the first exoplanet whose atmosphere was investigated using far-ultraviolet (far-UV) absorption line spectroscopy~\citep{vidal-madjar03,vidal-madjar04}.  The far-UV bandpass is unique in that 
it offers direct access to the strongest transitions of the atoms and molecules that constitute the majority of the mass in exoplanets (e.g. H, O, C, H$_{2}$, CO, etc.), and that are involved in the chemical reactions that produce more complex molecules observed in gas giants (H$_{2}$O, CH$_{4}$, CO$_{2}$, TiO, and VO;
Swain et al. 2009; Sing et al. 2009; D\'{e}sert et al. 2008).~\nocite{swain09,sing09,desert09}  Using the far-UV capabilities of $HST$/STIS,~\citet{vidal-madjar04}
detected absorption from \ion{H}{1} Ly$\alpha$, \ion{O}{1} 1304~\AA, and \ion{C}{2} 1335~\AA.  These results suggested that HD209458b has an extended atmosphere overfilling the Roche lobe and evaporating via a hydrodynamic escape mechanism.  We note however that data analysis techniques and corresponding interpretations can strongly influence conclusions drawn from these UV data~\citep{ben-jaffel07,vidal-madjar08}.  This situation is complicated for \ion{O}{1} and \ion{C}{2} by the low-resolution of the STIS data, and additional observations would be beneficial~\citep{linsky10}.

The majority of studies investigating the atmospheric conditions of transiting exoplanets have
employed absorption techniques.  In addition to the UV studies cited above, infrared (IR) observers
have used both spectroscopic~\citep{swain09} and photometric~\citep{desert09} methods to constrain the content of complex molecules in exoplanet atmospheres.  Studies of exoplanetary thermal emission 
in the mid-IR have shown that these objects can be observed directly at wavelengths where
the relative planet/star contrast is favorable~\citep{knutson08,knutson09a}.  
Analogously, both Jupiter and Saturn are strong far-UV emitters, well above any contribution from scattered sunlight at $\lambda$~$\lesssim$~1700~\AA~\citep{feldman93,gustin02,gustin09}.

\begin{deluxetable}{cccc}[b]
\tabletypesize{\footnotesize}
\tablecaption{HD209458 COS observing log. \label{cos_obs}}
\tablewidth{0pt}
\tablehead{
\colhead{Dataset} & \colhead{COS Mode} & \colhead{Orbital Phase}   
& \colhead{T$_{exp}$ (s)} 
}
\startdata	
lb4m01	& 	G160M 	& 	0.00		&	 4900	 \\
lb4m02	& 	G160M 	& 	0.25 	&	 7891	 \\
lb4m03	& 	G160M 	& 	0.50 	&	 4912	 \\
lb4m04	& 	G160M 	& 	0.75 	&	 7751	 \\
lb4m05	& 	G130M 	& 	0.00 	&	 4947	  \\
lb4m06	& 	G130M 	& 	0.25 	&	 7942	 \\
lb4m07	& 	G130M 	& 	0.50 	&	 4957  	 \\
lb4m08	& 	G130M 	& 	0.75 	&	 7796  	 \\
 \enddata

\end{deluxetable}

Far-UV emission from Jupiter and Saturn is dominated by atomic (Lyman series) and molecular hydrogen (H$_{2}$) lines, the primary constituents of gas giant planets~\citep{sudarsky03}.  
H$_{2}$ may be responsible for the Rayleigh scattering observed in optical spectra of the HD209458b atmosphere~\citep{lecavelier08}.
H$_{2}$ in the atmosphere of giant planets is excited by two primary mechanisms: electron bombardment and fluorescence pumped by stellar emission lines.  Both processes excite the ambient molecules to higher-lying electronic states whose decay produces a highly structured spectrum in the 700~--~1650~\AA\ bandpass.  The majority of these lines can be attributed to fluorescence from the Lyman and Werner bands of H$_{2}$ ($B$$^{1}\Sigma^{+}_{u}$~and~$C$$^{1}\Pi_{u}$~to~$X$$^{1}\Sigma^{+}_{g}$, however higher electronic states contribute to the observed electron impact spectrum).  Gas giant aurorae produce the highest surface brightnesses~\citep{clarke98} as the electron impact spectrum is most intense where the magnetic field lines connect with the atmosphere.  Dayglow (non-auroral illuminated disk) emission from the giant planets also shows strong features of H$_{2}$, with differing levels of molecular excitation and contribution from solar fluorescence~\citep{feldman93,wolven98}.  It follows that the far-UV could be an ideal wavelength regime in which to probe the atmospheres of hot Jupiters through their direct
emission~\citep{yelle04}.    Most transiting extrasolar giant planets are at relatively small separations from their parent star ($a$~$\lesssim$~0.1~AU), where the line flux from the host star should be more intense than at Jupiter by a large factor ($\gtrsim$~10$^{3}$) and bombardment from a stellar wind may be enhanced relative to the Jovian environment.  Additionally, by virtue of their tight orbits, these planets have large orbital velocities ($\gtrsim$~100 km s$^{-1}$).  Hence, by observing the system near both quadrature positions (Phase 0.25 and Phase 0.75), even with modest spectral resolution, the observed velocity shift between the two epochs can provide confirmation that the signal is planetary in origin.  With a velocity resolution of $\sim$~15 km s$^{-1}$ and an order of magnitude increase in sensitivity over previous far-UV instruments, the Cosmic Origins Spectrograph (COS) enables new opportunities for studies of exoplanetary atmospheres.

In this paper, we present far-UV observations designed to search for direct emission from an 
extrasolar planet.  We use early observations from COS, 
recently installed on $HST$, to constrain the emission properties of bulk atmospheric 
constituents such as H$_{2}$, as well as several atomic species.   We describe the COS observations and data reduction in Section 2.  In Section 3, we describe the details of an exoplanet auroral emission model based on observations of Jupiter, and present a quantitative analysis of the far-UV spectrum of HD209458(b) obtained with COS.  In Section 4, we compare the data to synthetic spectra, and place limits on the H$_{2}$ emission in the HD209458b atmosphere.  
We conclude with a brief summary in Section 5.

\begin{figure*}
\begin{center}
\hspace{+0.0in}
\epsfig{figure=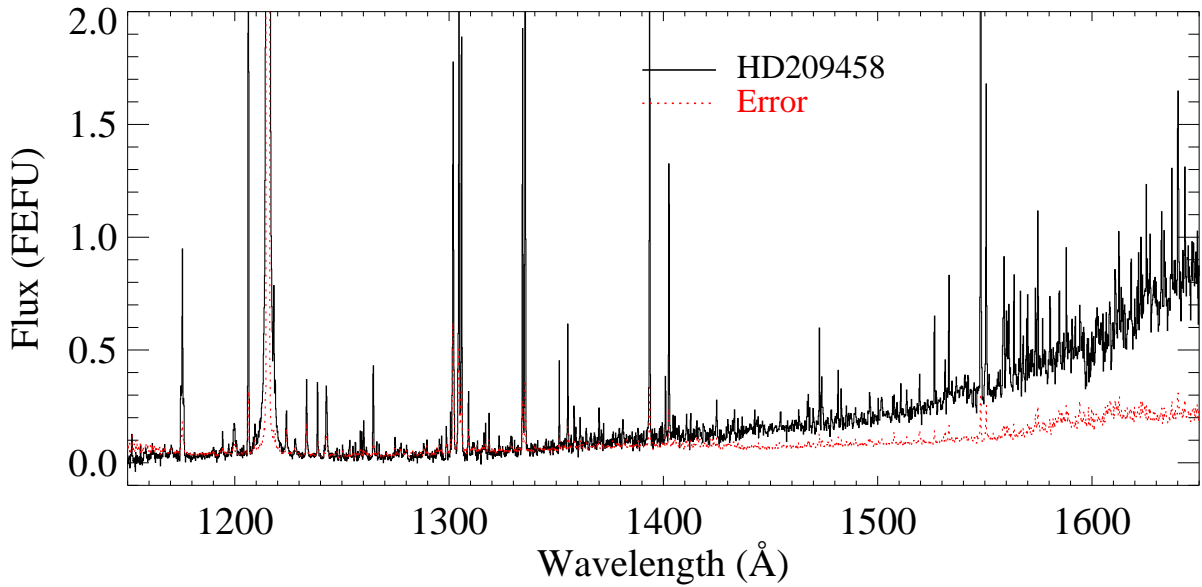,width=7.0in,angle=0}
\caption{\label{cosovly} The far-UV spectrum of HD209458.  This plot is a
weighted coaddition of Phases 0.25, 0.5, and 0.75, combining data from the G130M and G160M 
modes of $HST$/COS.  The flux units are those preferred in the COS instrument handbook, the femto-erg flux unit, FEFU (1 FEFU~=~1 $\times$~10$^{-15}$ ergs cm$^{-2}$ s$^{-1}$ \AA$^{-1}$).
 }
\end{center}
\end{figure*}

\section{$HST$-COS Observations and Data Reduction}

The Cosmic Origins Spectrograph is a fourth-generation instrument installed on 
$HST$ in 2009 May as part of Servicing Mission 4, STS-125.  
The spectrograph is optimized for observations of point sources
in the far-UV bandpass (1150~$\leq$~$\lambda$~$\leq$~1750~\AA) and achieves high sensitivity in the far-UV by using a single optical element for dispersion, focus, and correction of the spherical aberration of the $Hubble$ primary mirror. COS covers the 1150~--~1750~\AA\ band at moderate resolving power (G130M and G160M; $R$~$\simeq$~20,000; $\Delta$$v$~$\approx$~15 km s$^{-1}$ for point source observations), with additional modes that offer low resolution spectroscopy to wavelengths as short as the Lyman Limit (G140L; $R$~$\simeq$~2000; 900~$\lesssim$~$\lambda$~$\leq$~2100~\AA; McCandliss et al. 2009) as well as low and medium-resolution spectroscopy in the near-UV (G185M, G225M, G285M, and G230L; 1700~$\lesssim$~$\lambda$~$\leq$~3200~\AA).  First COS science results can be found in~\citet{france09} and~\citet{danforth09}, and a full instrument description and on-orbit 
performance characteristics are in preparation (J. C. Green et al. 2010, in preparation; S. Osterman et al. 2010, in preparation). ~\nocite{green10,osterman10,mccandliss09}

HD209458 was the first science target to be observed with COS after nominal science operations began in 2009 September (program 11534, $HST$ dataset {\tt lb4m}).  The observing program was divided into eight visits, four phases of the exoplanet orbit in G130M and G160M.  The relevant observational parameters are given in Table 1, and we used the orbital period and zero phase data from~\citet{knutson07}, $P$~=~3.52474859 days, $T_{C}$~=~2,452,826.628521, to determine the orbital phase during the observations.
  The transit duration for HD209458b is $\sim$~3.2 hrs, thus a single transit spans approximately two $HST$ orbits.  The transit data will be presented in Linsky et al. (2010), and we refer the reader to this work for a discussion of atomic absorption from the extended atmosphere of HD209458b.  In the present work, we will focus on the 
observations made at the Phase 0.25 and Phase 0.75 quadrature positions, using the Phase 0.5 secondary eclipse of the planet by the star (hereafter ``occultation'') observations as a baseline reference for any emission signal attributable to the planet.~\nocite{linsky10}  We present the 1150~--~1650~\AA\ spectrum in Figure 1.

HD209458 was observed by $HST$-COS in the interval from 15 September 2009 to 
18 October 2009, for a total of 20 orbits.  All observations were centered on the host star (R.A. = 22$^{\mathrm h}$ 03$^{\mathrm m}$ 10.82$^{\mathrm s}$, 
Dec. = +18\arcdeg 53\arcmin 03.8\arcsec ; J2000) and COS performed target acquisition using near-UV imaging through the Bright Object Aperture (BOA).  In order to achieve continuous spectral coverage and minimize fixed pattern noise, observations
in each grating were made at four central wavelength settings ($\lambda$1291,
1300, 1309, and 1318 in G130M and $\lambda$1589, 1600, 1611, and 1623 in G160M).
The data were processed with the COS calibration pipeline, CALCOS\footnote{We refer the reader to the cycle 18 COS Instrument Handbook for more details: {\tt http://www.stsci.edu/hst/cos/ documents/handbooks/current/cos\_cover.html}},
v2.11.  During the analysis, we observed anomalous behavior in several stellar emission lines that was traced back to a change in flux calibration reference files over the
course of the observing period.  The entire data set was reprocessed following 
completion of the program using the most recent\footnote{as of 2009 November 1} COS sensitivity files.

\section{HD209458b Auroral Model and Spectral Analysis} 

\subsection{Atmospheric H$_{2}$ Emission Model}

In this section, we present an H$_{2}$ model for the atmospheres of giant planets.  We describe the model
in general terms as a tool for future far-UV analyses of exoplanetary dayglow/aurorae, and use this model to 
compare with the COS observations in Section 4. 
Synthetic spectra of fluorescent emission from H$_{2}$ can 
be made by computing the collisional and radiative excitation rates into the
upper electronic states.  This model assumes a ground electronic
state population, then uses excitation cross sections and an incident radiation field to calculate the population distribution in the rovibrational levels of 
the upper electronic state (predominantly $B$$^{1}\Sigma^{+}_{u}$~and~$C$$^{1}\Pi_{u}$).
The molecules will then return to the ground electronic state following
the appropriate branching ratios, producing far-UV emission lines (700~$\leq$~$\lambda$~$\leq$~1650~\AA) and leaving the molecules in excited rovibrational levels.  We have adopted a modified version of the synthetic
molecular hydrogen emission code presented by~\citet{wolven97} to model fluorescence induced by 
electron impact and solar Ly$\alpha$ at the Shoemaker-Levy 9 impact site on Jupiter (see also models of the far-UV Jovian dayglow presented by Liu \& Dalgarno 1996).~\nocite{liu96}

We begin by creating an electronic ground state population for the molecules. This is done following the formalism of Herzberg to create a matrix of possible energy levels, characterized by their
rotational and vibrational quantum numbers, $J$ and $v$, respectively~\citep{herzberg}.  
An occupation probability of a given rovibrational state, $P(v,J)$, is found multiplying the Boltzmann factor by the degeneracy of a given state.  

The photon-induced fluorescence starts from the ground state population, the molecules are 
then ``illuminated'' by the EUV97 model solar radiation field 
($F_{\odot}$ from 1~--~2000~\AA; Tobiska \& Eparvier 1998).~\nocite{tobiska98}  
Excitation into the upper states 
is determined by theoretical rates at the Born limit, using the
transition probabilities of Abgrall et al. (1993a,b).  The excitation rate is given by
\begin{equation}
R(vJ \rightarrow nv'J')~=~I_{\lambda} \frac{\tau^{n}_{\lambda}}{\tau_{\lambda}}	\\
\end{equation}
where $v$ and $J$ are the vibrational and rotational 
quantum numbers of the ground electronic state and $v'$ and $J'$
are the vibrational and rotational quantum numbers of the 
electronic level $n$.  The units of $R$ and $I_{\lambda}$ are 
photons cm$^{-2}$ s$^{-1}$ \AA$^{-1}$.
$\tau_{\lambda}$ is the total optical depth 
at a given wavelength, and $\tau^{n}_{\lambda}$ is the optical 
depth for the transition connecting $vJ$ and $nv'J'$.  The intensity
of radiation being absorbed at a given wavelength, $I_{\lambda}$, is 
\begin{equation}
I_{\lambda}~=~ I^{o}_{\lambda}(1 - e^{-\tau_{\lambda}})	
\end{equation}
where $I^{o}$ is the model solar spectrum, with a scaling 
proportional to the HD209458~--~HD209458b distance.

Following the absorption, the molecules emit obeying branching ratios.
The emission rate from a particular level of $nv'J'$ to a 
level $v''J''$ of the ground electronic state is given by
\begin{equation}
S(nv'J' \rightarrow v''J'')~=~ R(nv'J')
[1 - \eta(nv'J')]\frac{A_{nv'J' \rightarrow v''J''}}{A_{nv'J'}}	
\end{equation}
where $R(nv'J')$ is the excitation rate of the excited electronic level (Equation 1), 
$\eta(nv'J')$ is the efficiency for predissociation in the 
excited electronic state, and the ratio of transition probabilities 
($A_{nvJ}$; Abgrall et al. 1993a,b) 
is the branching ratio~\citep{liu96}.~\nocite{abgrall93a,abgrall93b}

These models include absorption out of upper vibrational states ($v$~$>$~0),
and a first-order correction for self-absorption by H$_{2}$
at wavelengths shorter than 1120~\AA.  These models also take into account electronic transitions
to the $B^{'}$, $B^{''}$, $D$, and $D^{'}$ states, although their
relative contribution to the resultant spectrum is small.  Finally, 
transitions to predissociating states and vibrational states that 
result in dissociation ($v^{''}~>$~14, the vibrational continuum) are included.  

\begin{figure}
\begin{center}
\hspace{+0.0in}
\epsfig{figure=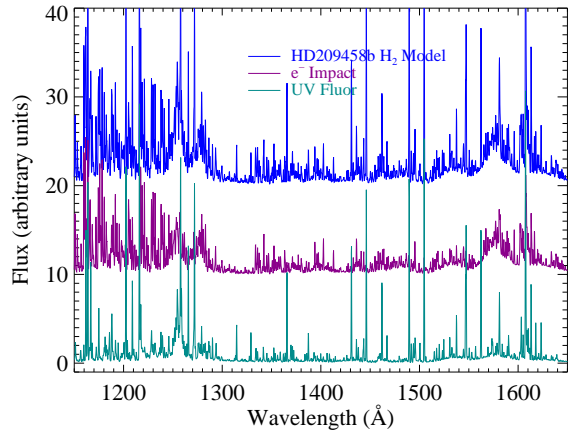,width=2.5in,angle=90}
\caption{\label{cosmod} Model H$_{2}$ fluorescence spectra, based on a model of the Jovian atmosphere
presented in~\citet{wolven97}.  This model was computed with $N(H_{2})$~=~10$^{20}$ cm$^{-2}$, 
$T(H_{2})$~=~1000 K, and $E_{e}$~=~100~eV.  The model is described in detail in \S3.1.  
Here we separate the total predicted model ($blue$) into components from electron impact-induced fluorescence ($dark~magenta$) and fluorescence excited by stellar emission lines ($dark$~$cyan$), notably
Ly$\alpha$, Ly$\beta$, \ion{C}{3}, and \ion{O}{6}. The model is rebinned to 0.5~\AA\ resolution for display purposes.}
\end{center}
\end{figure}

In addition to the incident radiation field, the model inputs are
the ground state thermal temperature ($T(H_{2})$), the column density ($N(H_{2})$), 
the electron energy distribution ($E_{e}$), the Doppler $b$-value (in km s$^{-1}$, 
contributing to the line shape), and a correction for attenuation by atomic
hydrogen on the incident stellar spectrum.  There is evidence that the H$_{2}$ temperature in the atmosphere 
of HD209458b is as high as 2350~K~\citep{lecavelier08}, however, the column density of the 
H$_{2}$ in the atmosphere of HD209458b is unknown, thus we use the Jovian system as a 
starting point for the creation of a synthetic auroral/dayglow spectrum.  We created a grid of 
models for several values of $T(H_{2})$, $N(H_{2})$, and $E_{e}$.  For the purposes of line
identification and H$_{2}$ detection, we employed a model with $T(H_{2})$~=~1000~K
(as there may be multiple interpretations for the data described in Lecavelier et al. 2008, we choose a $T(H_{2})$ in the range of the assumed effective temperature for a model exoplanet H$_{2}$ layer, 1000~--~1500 K; Yelle 2004; Murray-Clay et al. 2009), $N(H_{2})$~=~1~$\times$~10$^{20}$~cm$^{2}$~\citep{wolven97,wolven98}, and $E_{e}$~=~100 eV~\citep{liu96}.~\nocite{yelle04,murray-clay09}
The $b$-value is simply (2$kT(H_{2})/m)^{1/2}$, approximately 3 km s$^{-1}$ for the 
model spectrum described here, however the output spectrum is largely insensitive to the actual choice of $b$-value for values of a few km s$^{-1}$.  In the general case presented here, we neglect absorption by atmospheric neutral hydrogen, however this is most likely not proper for the HD209458b atmosphere, and we will return to this point in \S4.2.1.

We note that while these values most likely do not represent the exact conditions in the 
HD209458b atmosphere, the qualitative features of the output emission spectrum are 
relatively insensitive to the specifics for $N(H_{2})$~$<$~10$^{21}$~cm$^{2}$ and
$T(H_{2})$~$>$~300 K.  Very large column densities produce enhanced 
emission at long wavelengths ($\lambda$~$>$~1350~\AA) as the 912~--~1110~\AA\ 
flux gets self-absorbed 
and redistributed.  H$_{2}$ at high temperatures ($T(H_{2})$~$\gtrsim$~2500 K)
is characterized by strong band emission in the 1200~--~1280~\AA\ range because the broadly distributed ground state population allows enhanced absorption into the Werner bands which dominate the emission 
spectrum following the fluorescent cascade.  

\begin{deluxetable*}	{lcccc}
\tabletypesize{\small}
\tablecaption{HD209458 ultraviolet emission line catalog, line strengths, and line widths\tablenotemark{a}. \label{hd209458_lines}}
\tablewidth{3.0in}
\tablehead{
\colhead{Species} & \colhead{$\lambda_{rest}$} & \colhead{$\lambda_{obs}$} & \colhead{Line Flux}   
& \colhead{FWHM}   \\ 
    & (\AA) & (\AA) & (10$^{-16}$ ergs cm$^{-2}$ s$^{-1}$) & (km s$^{-1}$)  
}
\startdata
\ion{C}{3}\tablenotemark{b}  &      1176   &     1174.94 &   2.19 $\pm$  0.31 &        144.5 \\ 
\ion{C}{3}  &      1176   &     1175.56 &   4.10 $\pm$  0.67 &        106.7 \\ 
\ion{C}{3}  &      1176   &     1176.20 &   0.77 $\pm$  0.44 &        48.6 \\ 
\ion{N}{1}  &     1199.97   &     1199.84 &   1.00 $\pm$  0.30 &        136.3 \\ 
\ion{Si}{3}  &     1206.50   &     1206.34 &   12.89 $\pm$  0.35 &        73.0 \\ 

\ion{H}{1}\tablenotemark{c}$\oplus$ &      1215.67   &     1215.49 &   383 $\pm$  16.71 &        209.3 \\ 

\ion{O}{5}  &      1218.34   &     1218.21 &   3.97 $\pm$  0.15 &        115.3 \\ 
\ion{N}{5}  &      1238.82   &     1238.62 &   1.08 $\pm$  0.58  &        74.9 \\ 
\ion{N}{5}\tablenotemark{d}  &      1242.80   &     1242.50 &   0.26 $\pm$  0.04 &        19.2 \\ 
\ion{N}{5}\tablenotemark{d}  &      1242.80   &     1242.94 &   1.30 $\pm$  1.07  &        134.8 \\ 

\ion{S}{2}  &      1259.52   &     1259.08 &  1.68 $\pm$  0.54 &       170.5 \\ 
\ion{Si}{2}  &      1264.74   &     1264.71 &   2.11 $\pm$  0.11 &        117.4 \\ 
\ion{C}{1}  &     1274.98   &     1274.80 &   0.16 $\pm$  0.049 &        28.1 \\ 
\ion{Si}{3}  &      1298.95   &     1298.85 &   0.63 $\pm$  0.09 &        127.8 \\ 

\ion{O}{1}$\oplus$  &      1302.17   &     1301.89 &   7.38 $\pm$  0.62 &        70.4 \\ 
\ion{O}{1}$\oplus$  &      1304.86   &     1304.63 &   11.79 $\pm$ 0.17 &        133.0 \\ 
\ion{O}{1}$\oplus$  &      1306.03  &     1305.80 &   7.42 $\pm$  0.49 &        66.8 \\ 
\ion{Si}{2}  &      1309.28   &     1309.12 &   3.15 $\pm$  0.48 &        209.6 \\ 

\ion{N}{1}  &      1319.00   &     1318.71 &  0.28 $\pm$  0.06 &       21.5 \\ 
\ion{C}{2}  &      1334.53   &     1334.32 &   8.04 $\pm$  0.25 &        72.4 \\ 
\ion{C}{2}  &      1335.71   &     1335.50 &   16.06 $\pm$  0.14 &        70.2 \\ 
\ion{Cl}{1}  &     1351.66   &     1351.45 &   1.20 $\pm$  0.30 &        70.6 \\ 
\ion{O}{1}$\oplus$ &       1355.60   &     1355.39 &   1.38 $\pm$  0.09 &        48.7 \\ 

\ion{Si}{4}  &      1393.76   &     1393.58 &   9.47 $\pm$  0.40 &        70.1 \\ 
\ion{Si}{4}  &      1402.77   &     1402.61 &   5.11 $\pm$  0.49 &        70.8 \\ 

\ion{Fe}{2}  &      1418.53   &     1418.65 &   0.42 $\pm$  0.05 &        41.7 \\ 
\ion{S}{1}  &      1425.03   &     1424.89 &   0.63 $\pm$  0.05 &        29.4 \\ 
\ion{S}{1}  &      1433.28   &     1433.13 &   0.41 $\pm$  0.08 &        37.4 \\ 

\ion{C}{1}  &      1470.09  &     1469.97 &   0.36 $\pm$  0.08 &        36.0 \\ 
\ion{S}{1}  &      1472.97   &     1472.82 &   1.56 $\pm$  0.33 &        42.1 \\ 
\ion{S}{1}  &      1474.00   &     1473.89 &   1.33 $\pm$  0.33 &        34.6 \\ 
\ion{S}{1}  &      1481.70   &     1481.54 &   0.74 $\pm$  0.22 &        38.1 \\ 
\ion{S}{1}  &      1483.23   &     1483.00 &   0.31 $\pm$  0.04 &        35.8 \\ 

\ion{S}{2}\tablenotemark{e}  &      1519.91   &     1519.78 &   1.58 $\pm$  0.87 &        121.4 \\ 
\ion{Si}{2}  &     1526.71   &     1526.48 &   1.19 $\pm$  0.072 &        23.1 \\ 
\ion{C}{3}\tablenotemark{e}	  &       1531.84   &     1531.43 &   0.75 $\pm$  0.13 &        39.1 \\ 
\ion{Si}{2}  &     1533.43   &     1533.27 &   2.31 $\pm$  0.14  &        63.7 \\ 

\ion{C}{4}  &      1548.19   &     1548.05 &   10.54 $\pm$  0.38 &        72.1 \\ 
\ion{C}{4}  &      1550.77   &     1550.58 &   7.07 $\pm$  0.96  &        108.5 \\ 
\ion{Fe}{2}  &     1559.08   &     1558.93 &   1.98 $\pm$  0.29  &        61.1 \\ 
\ion{Fe}{2}  &     1563.79   &     1563.61 &   1.38 $\pm$  0.74  &        26.8 \\ 
\ion{Fe}{2}  &     1566.82   &     1566.63 &   0.73 $\pm$  0.67  &        16.0 \\ 
\ion{Fe}{2}  &     1570.24   &     1570.06 &   0.64 $\pm$  0.40  &        26.6 \\ 
\ion{Si}{1}+\ion{Fe}{2}  &      1573.87   &     1573.71 &   1.59 $\pm$  0.10 &        59.6 \\ 
\ion{Si}{1}+\ion{Fe}{2}  &      1574.82   &     1574.77 &   1.67 $\pm$  0.10 &        41.8 \\ 

\ion{Fe}{2}  &     1580.63   &     1580.38 &   1.22 $\pm$  0.21 &        42.4 \\ 
\ion{Fe}{2}  &     1584.95   &     1584.59 &   1.06 $\pm$  0.13 &        43.2 \\ 
\ion{Fe}{2}  &     1588.29   &     1588.09 &   1.23 $\pm$  0.13 &        33.9 \\ 

\ion{Fe}{2}  &     1612.80   &     1612.58 &   1.61 $\pm$  0.38 &        44.9 \\ 
\ion{Fe}{2}  &     1618.47   &     1618.29 &   0.74 $\pm$  0.10 &        58.6 \\ 
\ion{Fe}{2}  &      1621.68   &     1621.63 &   0.62 $\pm$  0.16 &        38.2 \\ 
\ion{Fe}{2}  &     1623.09   &     1622.88 &   0.65 $\pm$  0.06 &        22.9 \\ 
\ion{Fe}{2}  &     1625.52   &     1625.35 &   1.47 $\pm$  0.23 &        35.8 \\ 

\ion{Fe}{2}  &     1632.67   &     1632.51 &   1.66 $\pm$  0.55 &        68.2 \\ 
\ion{Fe}{1}  &     1637.40    &     1637.26 &   2.03 $\pm$  0.17 &        39.9 \\ 
\ion{He}{2}  &     1640.40   &     1640.15 &   2.51 $\pm$  0.38 &        63.2 \\ 
\ion{Fe}{2}  &     1643.58   &     1643.38 &   3.19 $\pm$  0.26 &        91.8 \\ 
\\
 \enddata

\tablenotetext{a}{~Line list created from an exposure time weighted coaddition of 
the quadrature and occultation spectra.} 
\tablenotetext{b}{Blending of the \ion{C}{3} multiplet led to three distinct fit
components.} 
\tablenotetext{c}{ 
~Lines labeled $\oplus$ are contaminated by geocoronal emission. } 
\tablenotetext{d}{Incomplete detector pulse-height screening by the CALCOS pipeline
leads to a spurious feature coincident with the \ion{N}{5} $\lambda$1242~\AA\ line.}
\tablenotetext{e}{Tentative identification, previously undetected.}
\end{deluxetable*}

Synthetic spectra are presented in Figure 2, where we show a decomposition of the predicted H$_{2}$
spectrum into electron impact and ultraviolet line fluorescence components.  There are several notable features of this spectrum in the COS (G130M + G160M) bandpass shown here:  
\begin{itemize}
\item Two prominent Lyman band systems roughly centered on 1164 and 1173~\AA\ 
\item Strong Werner band emission from 1200~--~1280~\AA\
\item The ``fluorescent H$_{2}$ desert'' from 1300~--~1350~\AA\ 
\item Several strong photon-excited emission lines in the 1350~--~1550 region (mainly excited by stellar Ly$\alpha$ and Ly$\beta$ photons)
\item Dissociation continuum emission observed between 1350~--~1630~\AA\ resulting from transitions to the unbound vibrational continuum ($v^{''}$~$>$~14) of the ground electronic state (Stecher \& Williams 1967; Abgrall et al. 1997; see also France et al. (2005) for additional discussion of this feature in the context of photo-excited gas)~\nocite{stecher67,abgrall97,france05a}
\item The strongest individual feature in this H$_{2}$ spectrum is the $B$~--~$X$ (6~--~13) P(1) emission 
at 1607.50~\AA\ which is excited by coincidence of the ground state transition $B$~--~$X$ (6~--~0) P(1), 
1025.93~\AA, with stellar Ly$\beta$
\end{itemize}

\subsection{Stellar Line Inventory}

The planetary emission signal from the HD209458 system is expected to be weak compared to the stronger stellar features present in the COS far-UV bandpass.  As the far-UV spectrum of H$_{2}$ can have hundreds of relatively strong emission lines that depend upon the specifics of the molecular excitation, 
it is critical to have a thorough catalog of the stellar emission features in the 1150~--~1650~\AA\
H$_{2}$ range covered by COS.  In addition, while high-lying ions originating in the chromosphere or transition region of HD209458 (such as \ion{Si}{4} and \ion{C}{4}) will most likely not be found in an exoplanetary environment, it is possible that intermediate ionization species (\ion{C}{2}, \ion{O}{1}, \ion{S}{1}, \ion{Si}{2}; Vidal-Madjar et al. 2004) could exist in the hot thermosphere of a short-period exoplanet~\citep{yelle04,garcia07}.  Therefore, we compared detailed studies of the far-UV spectra of $\alpha$~Cen and Arcturus   
with our COS observations of HD209458 to catalog the stellar features~\citep{pagano04,hinkle05}.
Figure 3 displays representative spectral windows from the G130M and G160M observations, with the H$_{2}$ emission model shown offset in blue.  In Table 2, we list the observed emission lines, 
including their observed wavelength, integrated brightness, and line width.  We note that 
stellar variability calls into question the validity of line fluxes measured from data acquired over the course of $\sim$ 1 month.  While we did find relatively good agreement between the peak line strengths at the quadrature and occultation pointings ($\pm$~$\approx$~20\%), nothing in the following analysis depends on the absolute calibration of the stellar emission line fluxes.  The line fluxes quoted in Table 2
would best be thought of in a relative sense.

\begin{figure}
\begin{center}
\hspace{+0.0in}
\epsfig{figure=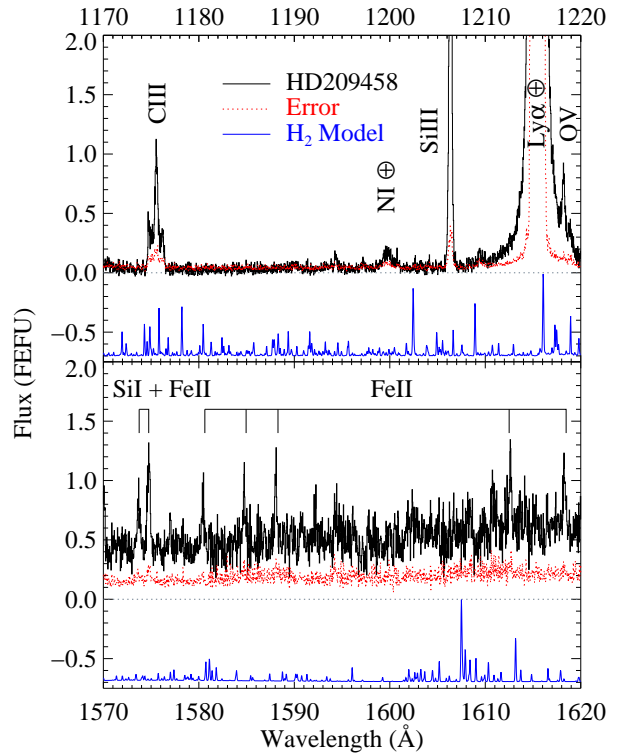,width=3.4in,angle=0}
\caption{\label{cosmod} Close-up views of individual spectral windows observed 
with COS G130M ($top$) and G160M ($bottom$).  Prominent stellar and geocoronal features
are labeled.  The model H$_{2}$ spectrum described in \S3.1 is shown offset in blue.
 }
\end{center}
\end{figure}

\subsection{Quadrature Observation Cross-Correlation}

Even in the far-UV bandpass, where continuum emission from solar type stars is weak (and consistent with the background at $\lambda$~$\lesssim$~1400~\AA), stellar contamination must be addressed.  
Steps must be taken in data analysis to maximize the potential emission from the planet while minimizing the contribution from stellar line and continuum emission.  We take advantage of the 
orbital velocity of HD209458b, $v_{orb}$~=~144 km s$^{-1}$ assuming a circular orbit, and coadd the 
quadrature observations as a function of velocity.  
Additionally, we subtract the stellar emission observed at the occultation (Phase 0.5) position in order to remove additional confusion from weak stellar lines.  Stellar variability is accounted for with a scaling based on the flux in high-ionization chromospheric/transition region lines (e.g. \ion{Si}{4} $\lambda$1400 and \ion{C}{4}~$\lambda$1550~\AA) not expected to be present in the planetary atmosphere.  We found that HD209458 was relatively stable across all of the observations used here (Section 3.1), with peak variability of $\Delta$$F($\ion{C}{4}$)$~$<$~5~$\times$~10$^{-16}$ ergs cm$^{-2}$ s$^{-1}$ \AA$^{-1}$, which is 
$\approx$ 19\% of the stellar flux at this wavelength.

This process not only increases the signal from the planet, but serves as a cross-correlation function that allows one to assess the validity of an emission signal.  We would expect any signal that originates on HD209458b to be at roughly the orbital velocity of the planet.  Mathematically, we define a star-subtracted quadrature spectrum ($D(v_{orb}$,$\lambda)$) which is a function of both wavelength and the orbital velocity of the planet, as 
\begin{equation}
D(v_{orb},\lambda)~=~( F_{Q1}(\lambda) + F_{Q2}(\lambda+d\lambda) ) - ( F_{Occ}(\lambda) + F_{Occ}(\lambda+d\lambda)	)  
 \end{equation}
with
\begin{equation}
d\lambda = \frac{2v_{orb}}{c}\lambda_{o}
 \end{equation}
and $\lambda_{o}$~=~1400~\AA, where $F_{Q1}$, $F_{Q2}$, and $F_{Occ}$ are the fluxed spectra (in units of ergs cm$^{-2}$ s$^{-1}$ \AA$^{-1}$) observed at the first quadrature (Phase 0.25), second quadrature (Phase 0.75), and occultation (Phase 0.5), respectively.  We create this planetary spectrum as a function of orbital velocity for the range 50~$\leq$~$v_{orb}$~$\leq$~250 km s$^{-1}$, in 1 km s$^{-1}$ bins.  In order to suppress noise spikes introduced during the subtraction, we rebin $D(v_{orb}$,$\lambda)$ to 0.1~\AA\ pixel$^{-1}$ ($\approx$~20 km s$^{-1}$), though numerous binnings were tested in order to rule out spurious features introduced by a particular choice of pixel scale.  The correlation signal in this treatment is the peak flux of $D(v_{orb}$,$\lambda)$
in a given velocity bin and spectral range, 
\begin{equation}
C(v_{orb})~=~MAX(D(v_{orb},\Delta\lambda)) 
\end{equation}
where $\Delta\lambda$ is the wavelength interval around the emission line of interest, typically $\sim$~10~\AA.  

Several spectral regions were identified to isolate specific emission lines of interest that fell into one of the following categories:
1) emission line-free, a region containing stellar continuum or detector background, 2) emission line regions containing only stellar features, 3) emission from low-ionization ions and neutrals which could have components from both the star and the planet, and 4) regions that should contain a wealth of planetary H$_{2}$ features.  A full list of the spectral windows investigated is provided in Table 3.  We then look for peaks in the correlation signal, $C(v_{orb})$, as a function of orbital velocity. 
Figure 4 displays the results for six of these regions.  We also measure the wavelength of peak emission in order to verify that the signal is associated with a specific line and not a chance noise peak.   As an illustrative comparison, we describe plots (a) and (b) in Figure 4.  These are regions dominated by stellar continuum (1500~--~1510~\AA) and stellar emission lines (\ion{C}{4} doublet, 1545~--~1555~\AA) respectively.  The left panels show the correlation peaks ($C(v_{orb})$, in units of 10$^{-15}$ ergs cm$^{-2}$ s$^{-1}$ \AA$^{-1}$)
as a function of orbital velocity.  Mean fluxes and standard deviations are computed from an average of two velocity regions distinct from the circular velocity (60~$\leq$~$v_{orb}$~$\leq$~100 and 200~$\leq$~$v_{orb}$~$\leq$~240 km s$^{-1}$) and displayed in Figure 4.
While the $C(v_{orb})$ plots show a mostly random distribution, the right panels show that while the peak wavelength of the continuum region is approximately randomly distributed (a), residuals in the fluxes in the \ion{C}{4} 1550.77~\AA\ are responsible for most of the peaks in the 1545~--~1555~\AA\ region (b).  

Figure 2 (c~--~f) shows four additional spectral regions that highlight possible emissions from the HD209458b system.  \citet{vidal-madjar04} 
observe possible absorption in the \ion{C}{2} 1335~\AA\ multiplet, arising in the extended envelope of the planet, therefore this could be a viable emitting species in a hot thermosphere~\citep{garcia07,linsky10}.  \ion{O}{1} 1356~\AA\
is considerably less contaminated by geocoronal emission than the \ion{O}{1} 1304~\AA\ multiplet, and could be present in the exoplanet.  In fact, the \ion{O}{1} 1356~\AA\ emission is observed to be brighter than \ion{O}{1} 1304~\AA\ in the Jovian system, where  \ion{O}{1} 1356~\AA\ is produced through the dissociation of O$_{2}$ and H$_{2}$O by electron impact~\citep{hall98,noren01}.
In panels (c) and (d), the peak wavelength, $\lambda_{C}$, displays a linear increase with orbital 
velocity, that we attribute to a broad geocoronal component that is being shifted to longer wavelengths in this treatment.
In the bottom panels (e and f), we display the correlation peaks for two regions with an abundance of strong H$_{2}$ features, 
centered around 1575 and 1608~\AA~\citep{gustin09}.  Nothing is seen in the 1608~\AA\ band, but there are several correlation peaks present in the 1575~\AA\ region within $\pm$~20 km s$^{-1}$ of the planet's nominal circular velocity.  The strongest two include a broad feature, approximately centered on the circular velocity of the planet, and the second is located at 158~--~160 km s$^{-1}$ ($\approx$~+15 km s$^{-1}$ in the rest frame of the planet).  We note that HD209458b is nearly circularized ($e$~=~0.014~$\pm$~0.009, consistent with a circular orbit; Laughlin et al. 2005), and the amplitude of variation in the orbital velocity is of order a few km s$^{-1}$.  The significance of these results and the possibility of an H$_{2}$ detection are discussed in \S4.

\begin{figure*}
\begin{center}
\hspace{+0.0in}
\epsfig{figure=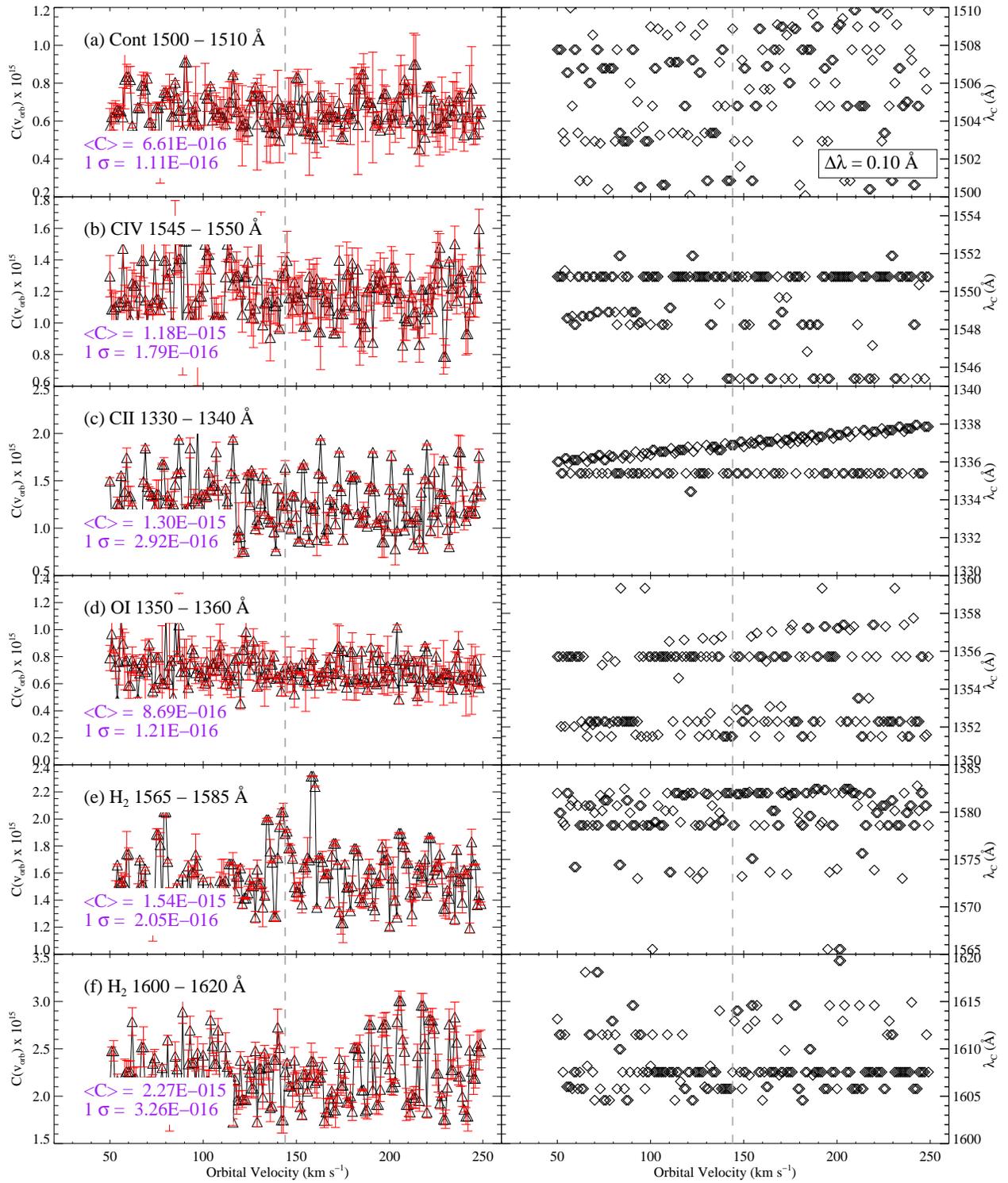,width=6.0in,angle=0}\vspace{0.3in}
\caption{\label{cosmod} Plots of the correlation peak, $C(v_{orb})$ (defined in \S3.3)
for orbital velocities from 50~--~250 km s$^{-1}$.  Several representative spectral regions are displayed:
(a) stellar continuum only, (b) stellar emission lines, (c) and (d) stellar atomic features with 
potential contributions from HD209458b, and (e) and (f) H$_{2}$ band regions.  
For each $v_{orb}$, the plots on the right show the wavelength corresponding to the maximum of $D(v_{orb}$,$\lambda)$.
The dashed gray vertical line represents the rest frame of the exoplanet for a circular velocity of 
144 km s$^{-1}$.
}
\end{center}
\end{figure*}

\section{Results and Discussion}

\subsection{Emission Feature at 1582~\AA\ }

An analysis of $C(v_{orb})$ for each of the spectral subregions identified in Table 3 indicates two statistically significant features that are consistent with emission from HD209458b.  There is 
a strong peak in the cross-correlation spectrum at $v_{orb}$~=~158~--~160 km s$^{-1}$ in the 
1575~\AA\ H$_{2}$ band. 
This 1582~\AA\ feature is detected at the 3.8 $\sigma$ level ($>$ 99.9\% confidence) at 1581.95~$\pm$~0.10~\AA\ (Figure 5).  The bottom panel of Figure 5 shows the quadrature summed, stellar subtracted spectrum ($D(v_{orb}$,$\lambda)$; Equation 4) for $v_{orb}$~=~159 km s$^{-1}$.  This feature is not obviously detected in either quadrature spectrum individually, and does not correspond to a known stellar feature.  In Figure 3, we showed that there are 
several \ion{Fe}{2} lines in this region.  The $HST$-STIS spectrum of $\alpha$ Cen~\citep{hinkle05} 
contains weak emission from \ion{Fe}{2} $\lambda$~1581.27~\AA.  While we cannot explicitly rule
out this line as the explanation for the 1582 \AA\ feature seen in HD209458b, we consider it unlikely that only this \ion{Fe}{2} line is present when dozens of stronger \ion{Fe}{2} lines are not observed.

We calculated the probability of randomly finding a 3.8~$\sigma$ event in the 166~\AA\ sampled in the 14 intervals (1660 total samples at 0.1~\AA\ binning) listed in Table~3.  Assuming Gaussian statistics, we find that one should expect 0.4 such events in the entire range, for each velocity interval.
This analysis indicates two (at $v_{orb}$~=~158 and 159 km s$^{-1}$) 3.8 $\sigma$ events, from which  
we estimate the possibility of a false positive at $\sim$~16\%.
It should be emphasized that this is the probability of finding a spurious signal anywhere in the 166~\AA\ sampled.  In reality, the false positive probability is much lower because there are only a small range of velocities that could be plausibly consistent with an exoplanetary origin.  Given the $\approx$~42 km s$^{-1}$ escape speed~\citep{linsky10}, a conservative range of possible atmpsheric emission velocities is $\pm$ 60 km s$^{-1}$, or $\pm$~$\sim$~0.6~\AA, using the binning described above.  This reduces the possibility of a random detection in each velocity interval by a factor of 20, leading to a realistic false positive probability of $\sim$~0.04\%. 
Thus, we conclude that this feature, at +15 km s$^{-1}$ in the rest frame of the planet, is most likely a real signal in the data.  

\begin{deluxetable}{cccc}[t]
\tabletypesize{\footnotesize}
\tablecaption{Quadtrature observation emission correlations. \label{cos_obs}}
\tablewidth{0pt}
\tablehead{
\colhead{Species} & \colhead{Origin\tablenotemark{a}} & \colhead{$\Delta\lambda$}   
& \colhead{Peak Flux Limit\tablenotemark{b}}  \\ 
 &   & \colhead{(\AA)}   
& \colhead{ } 
}
\startdata	
Background	& 	$\cdots$ 			& 	1145~--~1155		&	 1.34	 \\
\ion{C}{3}	& 	ST 				& 	1172~--~1180 		&	 1.30	 \\
\ion{H}{1}	& 	ST, PL, $\oplus$  	& 	1212~--~1222		&	 1.14	 \\
Background	& 	$\cdots$ 			& 	1280~--~1290 		&	 0.48	 \\
\ion{O}{1}	& 	ST, PL, $\oplus$  	& 	1300~--~1310		&	 2.56	 \\
\ion{C}{2}	& 	ST, PL 			& 	1330~--~1340 		&	 2.92	 \\
\ion{O}{1}	& 	ST, PL 			& 	1350~--~1360		&	 1.21	 \\
\ion{Si}{4}	& 	ST 				& 	1390~--~1406 		&	 2.32	 \\
\ion{S}{1}	& 	ST, PL 			& 	1470~--~1478		&	 0.99	 \\
Continuum		& 	ST 				& 	1500~--~1510 		&	 1.11	 \\
\ion{Si}{2}	& 	ST, PL 			& 	1524~--~1538		&	 1.31	 \\
\ion{C}{4}	& 	ST 				& 	1545~--~1555 		&	 1.79	 \\
H$_{2}$+\ion{Si}{1}+\ion{Fe}{2}		& 	PL, ST 	& 	1565~--~1585 	&	 2.05\tablenotemark{c}	 \\
H$_{2}$+\ion{Fe}{2}	& 	PL, ST 	& 	1600~--~1620	 	&	 3.26	 \\

 \enddata

\tablenotetext{a}{Proposed physical origin in the HD209458 observations: \\ST = Stellar, PL = Planet, 
$\oplus$ = Geocoronal airglow.} 
\tablenotetext{b}{in units of (10$^{-16}$ ergs cm$^{-2}$ s$^{-1}$ \AA$^{-1}$). 1 $\sigma$ limit on the orbital velocity cross-correlation, as described in 
\S3.3, computed over 60 $\leq$ $v_{orb}$ $\leq$ 100 and 200 $\leq$ $v_{orb}$ $\leq$ 240 km s$^{-1}$.} 
\tablenotetext{c}{Tentative H$_{2}$ detection at 3.8~$\sigma$ above this limit, discussed in \S4. }
\end{deluxetable}

There are several strong H$_{2}$ lines in this region, the brightest being
the Lyman band transitions ($B$~--~$X$) of (2~--~8) P(12),  (7~--~13) P(3),  (6~--~12) P(1), (2~--~9) P(4),
and (7~--~13) P(4), $\lambda_{rest}$~=~[1580.70, 1580.74, 1581.11, 1581.11, 1581.39~\AA], respectively.  
The oscillator strengths for each of the transitions are larger than 0.063, except for (6~--~12) P(1) which has $f$~=~0.027.  However, (6~--~12) P(1) has a pumping transition coincident with Ly$\beta$, as discussed in the following section.  The (7~--~13) lines are excited by electron impact, and are not seen elsewhere in the spectrum.  In the bottom panel of Figure 5, we overplot the UV photon fluorescence-only H$_{2}$ model in blue, shifted to $v_{orb}$~=~159 km s$^{-1}$.  The (2~--~8) P(12) line, which is pumped by coincidence with \ion{C}{3}~1175, is not seen.  Both of the 1581.11~\AA\ transitions are shifted to 1581.95~\AA\ at the velocity of the strong correlation peak.  The other features are displaced by $|\Delta\lambda|$~$\geq$~0.25~\AA, larger than the COS resolution element, and the 0.1~\AA\ binning applied to the $D(v_{orb}$,$\lambda)$ spectrum.  As noted in \S3.3, $D(v_{orb}$,$\lambda)$ was computed for several binnings to test for spurious features.  As would be expected for an unresolved molecular line, this feature is not detected for rebinning $\geq$~1~\AA.
We conclude that $B$~--~$X$ (6~--~12) P(1) $\lambda$1581.11~\AA\ and $B$~--~$X$ (2~--~9) P(4) $\lambda$1581.11~\AA\ are the most likely candidates for the possibly detected feature.  In Section 4.2 and 4.2.1, we will argue that this feature, if real, cannot be Ly$\beta$-pumped (6~--~12) P(1), and we describe a possible excitation mechanism for the (2~--~9) P(4) line.

\begin{figure}
\begin{center}
\hspace{+0.0in}
\epsfig{figure=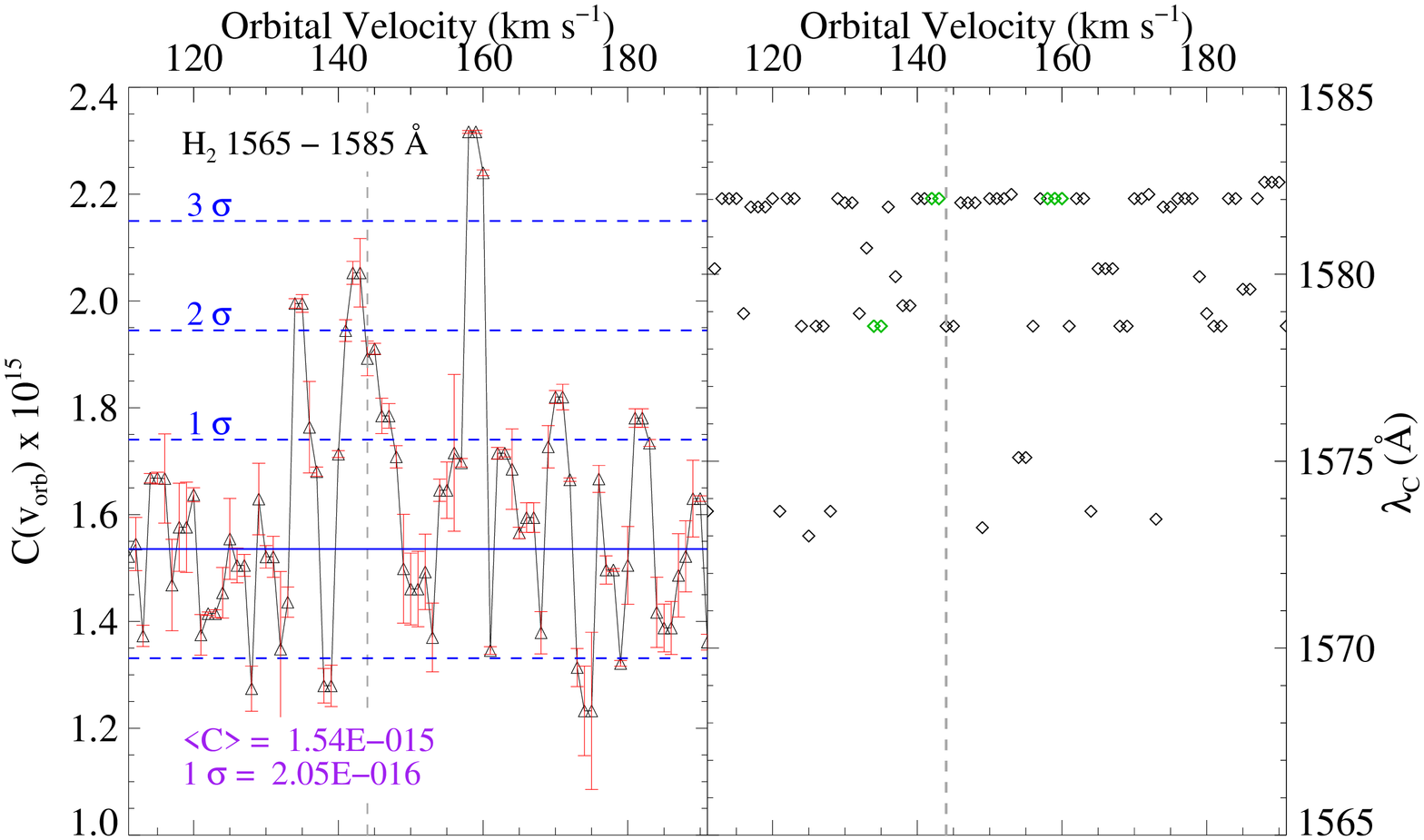,width=2.55in,angle=90}
\epsfig{figure=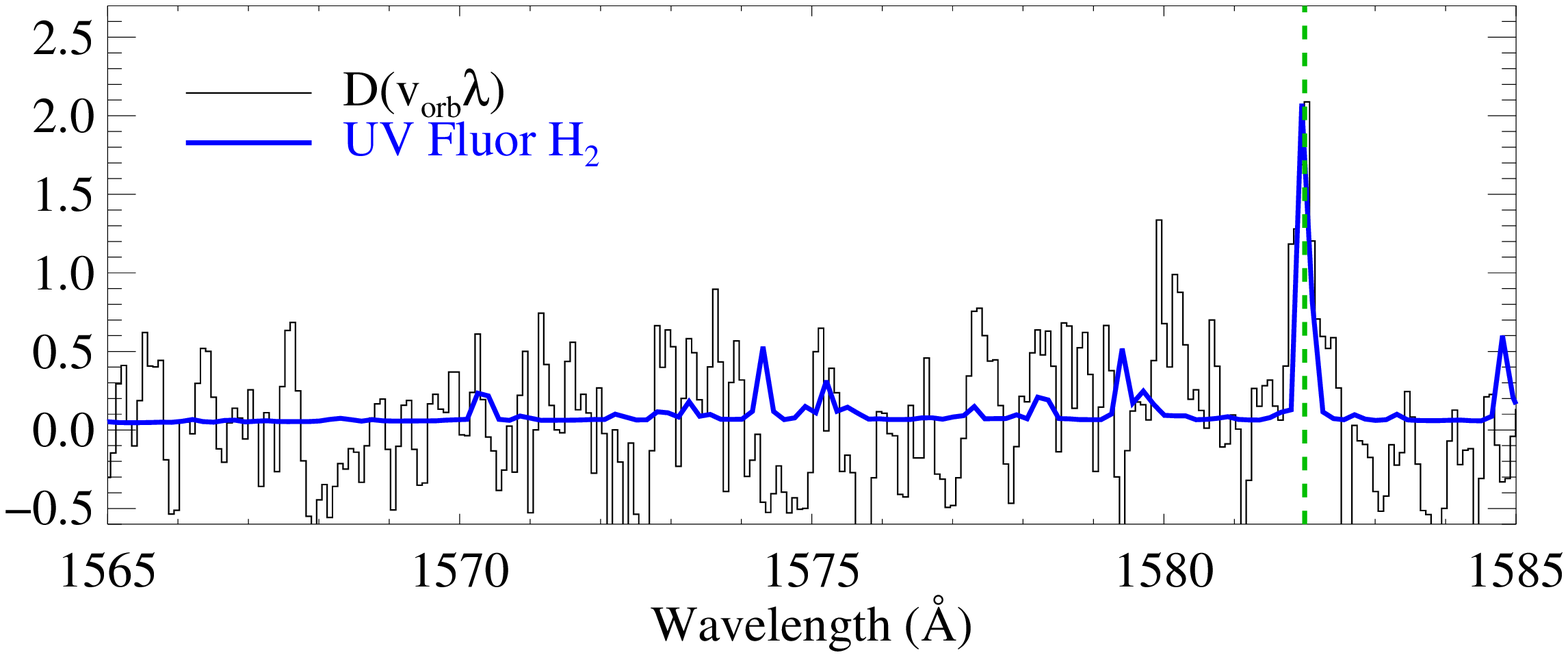,width=3.435in,angle=0}
\caption{\label{cosmod} Same as Figure 4, but showing the cross-correlation spectra in greater 
detail.  One observes three peaks found at greater than 2~$\sigma$ significance within
$\pm$ 20 km s$^{-1}$ of the planet rest frame ($v_{orb}$~=~144 km s$^{-1}$), and a peak 
at 158~--~160 km s$^{-1}$ detected at 3.8~$\sigma$ confidence.  The difference spectrum, 
$D(v_{orb}$,$\lambda)$, at $v_{orb}$~=~159 km s$^{-1}$  is shown in the lower panel, with the UV photon fluorescence model shown overplotted.  Both $B$~--~$X$ (2~--~9) P(4) and $B$~--~$X$ (6~--~12) P(1) emit at $\lambda_{rest}$
~=~1581.11~\AA, however, the absence of additional lines from the Ly$\beta$ progression (described in \S4.2.1) indicate that the observed line is $B$~--~$X$ (2~--~9) P(4), which may be excited by stellar \ion{O}{1} emission (\S4.2).
 }
\end{center}
\end{figure}

\begin{deluxetable*}{lcccc}
\tabletypesize{\footnotesize}
\tablecaption{Limits on H$_{2}$ emission from the atmosphere of HD209458b. \label{exo_lines}}
\tablewidth{0pt}
\tablehead{
\colhead{Line ID\tablenotemark{b}} & \colhead{$\lambda_{rest}$} & \colhead{Excitation} & \colhead{Line Flux}   
  &  \colhead{Notes\tablenotemark{c}} \\ 
    & (\AA) & & (10$^{-17}$ ergs cm$^{-2}$ s$^{-1}$ ) &   }
\startdata
$C$~--~$X$ (1~--~4) P(5)	& 	1171.96 	& e$^{-}$ Impact &	$\leq$  0.76 	&	 $\cdots$	 \\
$C$~--~$X$ (2~--~5) Q(1)	& 	1175.83 	& e$^{-}$ Impact &	$\leq$  6.22 	&	Blend with \ion{C}{3}  \\
$C$~--~$X$ (4~--~8) Q(1)	& 	1239.53 	& e$^{-}$ Impact &	$\leq$  0.44  	&	  $\cdots$		  \\
$B$~--~$X$ (1~--~3) R(5)	&   	1265.67   & UV photon, Ly$\alpha$	  &	$\leq$  0.71 	&	  $\cdots$	  \\
$C$~--~$X$ (7~--~12) Q(3)	& 	1279.10 	& e$^{-}$ Impact &	$\leq$  0.80 	&	  $\cdots$		  \\
$B$~--~$X$ (6~--~7) P(1)	& 	1365.66 	& UV photon, Ly$\beta$     &	$\leq$  2.01 	&	  $\cdots$		  \\
$B$~--~$X$ (6~--~9) P(1)	& 	1461.97 	& UV photon, Ly$\beta$     &	$\leq$  3.36 	&	 Stellar continuum at $\lambda$~$\gtrsim$~1420~\AA\	  \\
$B$~--~$X$ (1~--~8) R(3)	& 	1547.34 	& UV photon, Ly$\alpha$    &	$\leq$  5.83 	&	  $\cdots$		  \\
$B$~--~$X$ (7~--~13) P(3)	& 	1580.74 	& e$^{-}$ Impact &	$\leq$  10.14 	&	  Blend with \ion{Fe}{2} 		  \\
$B$~--~$X$ (2~--~9) P(4)	& 	1581.11 	& UV photon  &	11.72$\pm$3.08 	&	   Blend with \ion{Fe}{2}	   \\
$B$~--~$X$ (6~--~12) P(1)	& 	1581.11 	& UV photon, Ly$\beta$  &	$\cdots$ 	&	   Blend with \ion{Fe}{2}	   \\

$B$~--~$X$ (6~--~13) P(1)	& 	1607.50 	& UV photon, Ly$\beta$   &	$\leq$	   10.54 	&	  $\cdots$		  \\
$B$~--~$X$ (5~--~12) P(3)	& 	1613.18 	& e$^{-}$ Impact + &	 $\leq$	 12.80 	&	  $\cdots$	  \\
                                          	& 	& UV Photon, \ion{O}{6} &		  	&	  	  \\
 \enddata


\end{deluxetable*}

\subsection{H$_{2}$ Emission from HD209458b}

Strong auroral and/or dayglow H$_{2}$ emission from HD209458b is not observed.  
Figure 3 is representative of the results from comparisons between our COS observations and the models described in 3.1. 
Upper limits on the line strengths of the strongest emission lines predicted by our model are presented in Table 4.
As discussed above, the two candidate emission lines for the observed feature in the $D(v_{orb}$,$\lambda)$ spectrum are (6~--~12) P(1) and (2~--~9) P(4).  This feature peaks at +15 km s$^{-1}$ in the rest frame of the planet, with a second broader maximum approximately at rest in the planet frame (Figure 5).
The (2~--~9) P(4) line is intrinsically stronger than the (6~--~12) P(1), however a lack of clear excitation mechanism complicates this interpretation of the result.  There is no obvious pumping transition ((2~--~$v$) R(2) or P(4)) exactly coincident with an observed stellar emission line.  Possibilities exist if there is a velocity offset between the ``H$_{2}$ layer'' of the planet (1.0~$\leq$~$r$~$\leq$~1.1~$R_{P}$, where $R_{P}$ is the radius of HD209458b; Murray-Clay et al. 2009).~\nocite{murray-clay09}  For example, the (2~--~0) band is near strong \ion{N}{2} and \ion{He}{2} transitions ($\lambda$~1083~--~1086~\AA) observed in solar-type stars~\citep{redfield02}.  We analyzed $FUSE$ observations of HD209458 in order to constrain the stellar spectrum in this wavelength region, but the data were not of sufficient quality to derive meaningful constraints.  Another possibility is coincidence between (2~--~4) P(4) 1301.70~\AA\ and \ion{O}{1} 1302.  For $\delta$-function emission and absorption profiles, these lines would require a velocity offset of 108 km s$^{-1}$ to overlap, however in practice these lines will have an appreciable velocity width that may provide sufficient overlap~\citep{ben-jaffel09}.  

Exploring the possibility of \ion{O}{1} excitation further, we examined the relative line fluxes of stellar Ly$\alpha$ and \ion{O}{1}.  The data set for HD209458 is complicated by observations with different instruments, spectral resolutions, and the attenuation of both lines by 47 pc of ISM.  A somewhat more straightforward determination can be made using the high-ressolution STIS spectrum of $\alpha$ Cen~\citep{pagano04}.  Interstellar absorption still complicates the comparison, but we estimate an $F$(Ly$\alpha$)/$F$(\ion{O}{1}) ratio of $\sim$~20~--~40 for HD209458.
When considering the possibility of H$_{2}$ excited by stellar \ion{O}{1}, it is important to note that appreciable occupation of the $v$~=~4 level of the ground electronic state requires an energetic population of H$_{2}$, but this may be reasonable to expect in the atmosphere of a hot Jupiter.  The atmospheric temperature in the 
HD209458b atmosphere is expected to rise sharply at $r$~$>$~$R_{P}$~\citep{sing08b}, and this could increase the rovibrational temperature of the molecules in the H$_{2}$ layer with height, up to the collisional dissociation threshold for H$_{2}$ ($T$~$\approx$~4000 K; Shull \& Beckwith, 1982).~\nocite{shull82}
Additionally, there is evidence that the vibrational temperature of H$_{2}$ in the Jovian atmosphere can exceed the kinetic temperature by as much as 50\%~\citep{barthelemy05} and electronic excitation of 
H$_{2}$ from  highly excited levels of the ground state is observed in the warm H$_{2}$ environment around the classical T Tauri star TW Hya~\citep{herczeg02}.

\subsubsection{Constraints on Surface \ion{H}{1} Lyman Series Strength and Magnetic Field} 

Given the uncertainties in the excitation route for the (2~--~9) P(4) emission line described above, it would be tempting to assign the 1582~\AA\ emission to (6~--~12) P(1) pumped by stellar Ly$\beta$ through the (6~--~0) P(1) 1025.93~\AA\ transition.  This scenario does not hold as the (6~--12) band of the Ly$\beta$ pumped progression is among the weaker branches of (6~--~$v^{''}$).  There are several other lines excited by stellar Ly$\beta$ that should be $>$ 2 times brighter (and thus detectable), including the (6~--~13) P(1) $\lambda$1607.50~\AA\ line that should be $\sim$~4 times brighter than (6~--~12) P(1).  As none of these lines are observed, we are forced to rule out Ly$\beta$ pumping for the observed feature.  

The \ion{H}{1} Lyman series lines are the strongest far-UV spectral features in a solar-type star, and assuming that the 1582~\AA\ feature is produced by (2~--~9) P(4), this raises the question of why H$_{2}$ lines excited by Lyman photons are not observed.  \citet{vidal-madjar03,vidal-madjar04} describe the detection of an extended \ion{H}{1} atmosphere surrounding the planet, and we hypothesize that this warm cloud of neutral hydrogen is primarily responsible for the suppression of Lyman-pumped H$_{2}$ emission from HD209458b.  This \ion{H}{1} layer will scatter the incident Lyman series lines, effectively diffusing the 
photons away from the line core where the H$_{2}$ coincidence exists.
By comparing model line-strength ratios with the observed $F(\lambda_{rest}$1581.11 \AA)/$F(\lambda_{rest}$1547.34 \AA) and 
$F(\lambda_{rest}$1581.11 \AA)/$F(\lambda_{rest}$1607.50 \AA) ratios, we can constrain the relative optical depths seen by stellar Ly$\alpha$ and Ly$\beta$ photons before they reach the H$_{2}$ layer.  
We take the 1 $\sigma$ limits for the ``measured'' fluxes of the 1547.34 and 1607.50~\AA\ lines, and 
the model line ratios are $\approx$ 0.47 and 0.25, respectively.  
Assuming that attenuation of the pumping transition for (2~--~9) P(4) is negligible, we estimate that 
optical depths at the (1~--~2) R(6) 1215.73~\AA\ and (6~--~0) P(1) 1025.93~\AA\ transitions are 8.1 and 15.2, respectively.  We created a simulated \ion{H}{1} profile to determine the column density required 
to attenuate the stellar Lyman lines, finding $N(HI)$~$\gtrsim$~3~$\times$~10$^{20}$ cm$^{-2}$ on the line of sight from HD209458 to the H$_{2}$ layer on HD209458b.

The far-UV spectra of Jupiter and Saturn are dominated by electron impact excitation~\citep{feldman93,liu96,gustin09}, and in addition to understanding the lack of H$_{2}$ emission pumped by Lyman series photons, the apparent lack of electron impact H$_{2}$ must be addressed.  The energetic electron spectrum is driven by the interaction of the planetary magnetic field with the ambient atmosphere (Broadfoot et al. 1979, and see Bhardwaj \& Gladstone 2000 for a comprehensive review of the production of H$_{2}$ emissions in the Jovian aurorae).~\nocite{broadfoot79,bhardwaj00}  Thus, assuming that the atmosphere of HD209458b has an H$_{2}$-dominated composition similar to that of Jupiter, the paucity of electron impact emission can be interpreted as a reflection of a relatively weak surface magnetic field on HD209458b.  This interpretation is supported by theoretical studies of the magnetic environment of hot Jupiters.  The magnetic moment ($\mathcal{M}$) is known to scale with the core density of the planet, the rotation rate, and core radius~\citep{griesmeier04}.  A short-period exoplanet such as HD209458b is expected to be tidally locked due to gravitational dissipation, which results in a dramatically lower rotation rate compared to a gas giant such as Jupiter.  Combining the rotation rate, the density of the planet inferred from transit observations, and models of the interior structure, HD209458b is expected to have an intrinsic magnetic moment (in the tidally locked limit) of $\mathcal{M}$~$<$~0.01 $\mathcal{M}_{J}$~\citep{griesmeier04}.  
When factoring in the interaction with the stellar magnetic field, this value is of order 0.1~$\mathcal{M}_{J}$, consistent with the surface magnetic field calculations of~\citet{sanchez04}, $B$~$\leq$~0.1~$B_{J}$ (where $B_{J}$ is the Jovian polar magnetic field).  Interestingly, recent results from radio-frequency observations of the well-studied hot Jupiter HD189733b~\citep{lecavelier09} support the hypothesis of a weak  exoplanet magnetic field.
Another possible source of energy for atmospheric electrons is the interplanetary magnetic field ($B_{IP}$), but this is estimated to be small for the HD209458 system $B_{IP}$~$<$~5~$\times$~10$^{-3}$ $B_{J}$~\citep{erkaev05}.  We conclude that the feature observed at 1582~\AA\ in HD209458b is consistent with no detectable lines from electron impact H$_{2}$.
The slow rotation and low density of the planet creates a magnetic field that is incapable of producing H$_{2}$ emission with intensities observed in the Jovian system.


\subsubsection{Alternative Excitation Routes and H$_{2}$ Radiative Transfer Effects}

We have shown that the primary excitation routes for populating the Lyman and Werner bands
in gas giants in our solar system may be suppressed in the atmosphere of HD209458b.  It is worth noting that the higher lying electronic states of H$_{2}$ can be populated via electron impact and absorption of extreme UV (EUV; $\lambda$~$\leq$~911~\AA) photons.  Electron impact excitation of the $E$ and $F$ states of H$_{2}$ can populate the Lyman bands via a radiative cascade process ($E,F$~$^{1}\Sigma^{+}_{g}$~$\rightarrow$~$B$$^{1}\Sigma^{+}_{u}$ + $h\nu$), contributing as much as $\sim$~25\% of the total flux from the $B$~--~$X$ band emission~\citep{ajello82}.  We note that this process concentrates the emitted photons in the 1350~--~1500~\AA\ region, and does not favor the putative $B$~--~$X$ (2~--~9) P(4) emission feature.  Coupling this with the arguments against electron impact excitation noted above, we dismiss this as a possible energy source.  What we cannot rule out is the possibility that one of the higher-lying electronic states is photo-excited by EUV stellar line photons.  This cascade could preferentially populate a specific level of the $B$ state, causing the resultant emission spectrum to differ from that expected from direct excitation of the Lyman and Werner bands.  However, the upper \ion{H}{1} atmosphere of HD209458b should effectively attenuate any stellar EUV emission capable of exciting H$_{2}$, making this route unlikely.

One possibility that could explain the lack of H$_{2}$ features seen in the spectrum of HD209458b is the coherent scattering of pumping photons absorbed in the wings of the H$_{2}$ line profiles.  This partial frequency redistribution process~\citep{lupu09,barthelemy04} requires a more sophisticated
radiative transfer treatment than the synthetic H$_{2}$ models presented in this work, where all photons 
absorbed in a given transition are redistributed to the line core prior to emission.~\nocite{lupu09}
This partial redistribution process, which has been observed in the spectra of Jupiter and Saturn, acts to suppress the line-to-continuum ratio and decrease the contribution from subordinate transitions in a given progression.
Additional radiative transfer investigations into EUV excitation of H$_{2}$ and frequency redistribution would be very useful for the interpretation of future experiments characterizing the far-UV spectra of extrasolar giant planets.

\section{Summary}
We have presented new $HST$-COS observations aimed at characterizing aurorae on the extrasolar planet HD209458b.   We use a model for the auroral and dayglow spectra of Jupiter to predict the expected features and use this spectrum to guide the data analysis.  These data reveal no unambiguously identifiable signature of atomic or H$_{2}$ emission from the planet.  We do detect a significant feature at 1582~\AA, observed at +15 ($\pm$ 20) km s$^{-1}$ in the rest frame of the planet.  This feature may be the H$_{2}$ $B$~--~$X$ (2~--~9) P(4) line at $\lambda_{rest}$~=~1581.11~\AA, however the excitation mechanism for this line is unclear.  Based on this feature, we placed limits on the level of emission from electron impact and stellar fluorescent excitation.  We put these limits in the context of the HD209458 system, deriving a limit on the atmospheric column of \ion{H}{1}.  The non-detection of electron impact H$_{2}$ is consistent with predictions of a weak magnetic field due to tidal locking and the low density of the planet.  We suggest that future ultraviolet studies of auroral/dayglow emission from extrasolar giant planets would benefit by targeting the nearest transiting planet systems that do not show evidence for extended atomic and ionic atmospheres.

\acknowledgments
It is a pleasure to thank members of the COS team for valuable input, notably Brian Keeney, Eric Burgh, Dennis Ebbets, and Sally Heap.  In addition, K.F. appreciates discussions with Tom Ayres and Roxana Lupu regarding the far-UV properties of cool stars and H$_{2}$, and acknowledges encouragement from the IGwAD consortium during a portion of this work.
This work was support by NASA grants NNX08AC146 and NAS5-98043 to the University of Colorado at Boulder.

\bibliography{ms_apj}










\end{document}